\documentclass[%
 aip,
 jmp,%
 amsmath,amssymb,
 preprint,%
]{revtex4-1}

\usepackage{graphicx}
\usepackage{dcolumn}
\usepackage{bm}
\usepackage{amssymb}
\usepackage{graphicx}
\usepackage{subfigure}
\usepackage[centerlast]{caption2}
\usepackage{overpic}
\usepackage{setspace}
\usepackage{ulem}
\usepackage{amsfonts}
\usepackage{mathrsfs}
\usepackage{float}
\usepackage{amsmath}
\usepackage{graphicx}
\usepackage{threeparttable}
\usepackage{amsthm}
\usepackage{longtable}
\usepackage{enumerate}
\usepackage{array}
\usepackage{color}
\usepackage{threeparttable}
\usepackage{url}
\usepackage{lineno}

\usepackage{geometry}
\usepackage[colorlinks,linkcolor=blue,urlcolor=blue,anchorcolor=blue,citecolor=blue]{hyperref}
\newcommand{\mm}[1]{\mathrm{#1}}

\makeatletter
\renewcommand{\@thesubfigure}{\hskip\subfiglabelskip}
\makeatother

\makeatletter
\def\hlinew#1{%
  \noalign{\ifnum0=`}\fi\hrule \@height #1 \futurelet
   \reserved@a\@xhline}
\makeatother
\makeatletter

\newcommand{\Rmnum}[1]{\expandafter\@slowromancap\romannumeral #1@}
\makeatother

\makeatletter
\def\ExtendSymbol#1#2#3#4#5{\ext@arrow 0099{\arrowfill@#1#2#3}{#4}{#5}}
\def\RightExtendSymbol#1#2#3#4#5{\ext@arrow 0359{\arrowfill@#1#2#3}{#4}{#5}}
\def\LeftExtendSymbol#1#2#3#4#5{\ext@arrow 6095{\arrowfill@#1#2#3}{#4}{#5}}
\makeatother

\begin{document}

\preprint{AIP/123-QED}

\title[Submit to XXX]{Coupling of homogeneous and heterogeneous melting kinetics in polycrystalline materials}

\author{Meizhen Xiang}
\email{xiang\_meizhen@iapcm.ac.cn}
\affiliation{
Laboratory of Computational Physics, Institute of Applied Physics and Computational Mathematics, Beijing, 100088, China.
 \\
}%

\author{Yi Liao}
\affiliation{
School of Mechanical Engineering, Southwest Petroleum University, Chengdu, 610500, China.
}%

\author{Guomeng Li}%
\affiliation{
School of Mechatronical Engineering, Beijing Institute of Technology, Beijing 100081, China.
}%

\author{Jun Chen}%
 \email{jun\_chen@iapcm.ac.cn}
\affiliation{
Laboratory of Computational Physics, Institute of Applied Physics and Computational Mathematics, Beijing, 100088, China.
 \\
}%

\date{\today}

\begin{abstract}
Melting kinetics of polycrystalline materials is analyzed on the basis of a new model which explicitly couples homogeneous and heterogeneous melting mechanisms.
The distinct feature of this approach lies in its ability to evaluate
not only grain-size-distribution effects on the overall melting kinetics but also competitions between the two melting mechanisms.
For the first time, we reveal the three-part structure of temperature-time-transformation diagrams for melting of polycrystalline materials, through which it is
 possible to determine a critical temperature across which the dominant melting mechanism switches.
The critical temperature increases as the mean-grain-diameter decreases following a negative power-law. The results are qualitatively consistent with  experimental observations.
\end{abstract}

\keywords{Suggested keywords}
\maketitle

\section{Introduction}
Melting is one of the most important phase
transformations in materials science and engineering.
There are
commonly two mechanisms of melting: homogeneous melting and heterogeneous melting.
 The former is related to randomly nucleation and growth of
the daughter liquid phase inside bulk crystals \cite{Mattias2005How,Siwick2003An,Jin2001Melting}, and the latter is characterized by motion of melting fronts initiated from
grain boundaries (GBs) and material interfaces \cite{Tsao1986Asymmetric,Ivanov2007Kinetic}.
Kinetics corresponding to these two mechanisms have been treated separately in the literature.
On one hand, homogeneous melting kinetics is commonly described
by the Kolmogorov-Johnson-Mehl-Avrami (KJMA) model \cite{Kolmogorov1937Static,Johnson1939Reaction,Avrami1940Kinetics} and its extensions \cite{Kooi2004Monte,Farjas2006Modification,Tomellini2014Kolmogorov}.
On the other hand,
heterogeneous melting kinetics is investigated by modeling the melting front velocity
as a function of temperature, on the basis of thermodynamic approaches\cite{Mendelev2010Molecular}, diffusion theory\cite{Jackson2002Interface},
thermal collision theory\cite{Broughton1982Crystallization} or molecular dynamics simulations\cite{Ivanov2007Kinetic,Mazhukin2017Kinetic}.

Previous melting kinetics models commonly involve only one of the two mechanisms while ignore the other, focusing on an isolated single-crystal grain or an isolated
solid-liquid interface.
However, in practical scenarios, polycrystalline materials contain amounts of grains and GBs with various shapes and sizes, and homogeneous and heterogeneous
melting would take places simultaneously.
The coupling effects between the two melting mechanisms in polycrystalline materials is still poorly understood. For example,
what is the demarcation temperature at which the contribution of homogeneous melting overweighs
that of heterogeneous melting?  What is the temperature range in which the two melting mechanisms make comparable contributions?
These questions are not well addressed due to a lack of suitable melting kinetics models.
%
%

Here, we propose a new kinetics model
which explicitly couples the contributions of both the two melting mechanisms to better understand melting kinetics of polycrystalline materials.
The model allows to investigate effects of grain-size-distribution (GSD) on the overall melting kinetics.
More importantly, we show that through the three-part structure of the temperature-time-transformation (TTT) diagrams,  one can clearly reveal competitions between the two melting mechanisms.
It leads to prediction of a critical demarcation temperature at which the dominant melting mechanism switches. The dependence of the critical temperature on mean-grain-diameter (MGD) is calculated.

\section{Modeling equations}
Polycrystalline materials comprise of amounts of grains with various shapes and sizes.
Firstly, we concern the melting kinetics of an isolated grain.
To simplify modeling, we approximate the melting kinetics of an isolated polyhedral grain by that of a sphere grain
with the same volume.
For a sphere grain, we assume that the heterogeneous melting front is a spherical surface
shrinking from the outer GB towards the center.
During dynamical melting, the grain is divided into two domains: an outer hollow sphere and an inner sphere.
The outer hollow sphere represents the domain which has been swept by the moving melting front initiated from the GB
and is fully melted.
The inner sphere is partially melted due to randomly nucleation and growth of liquid phase (homogeneous melting). Based on this scenario, the melting kinetics
of an isolated grain with grain diameter $D$ is represented by
\begin{equation}\label{eq:CHH}
{\eta}(t;D)= 1 - \frac{d(t;D)^3}{D^3}\Big(1-\chi(t)\Big)
\end{equation}
where ${\eta}(t;D)$ is the overall liquid fraction of the grain at time $t$, $d$ is the diameter of the inner sphere,
$\chi$ is the liquid fraction in the inner sphere.
The heterogeneous melting kinetics is modeled as
$\dot{d} = -2U_{\mm{front}}$
where $U_{\mm{front}}$ is the melting front velocity, which is a function of temperature in the following form \cite{Mazhukin2016Temperature,Mazhukin2017Kinetic},
$U_{\mm{front}} =
a_{\mm{U}} \sqrt{\frac{3k_{\mm{B}}T}{M_{\mm{atom}}}}
\Big[\exp\Big(b_{\mm{U}}\frac{\Delta H_{\mm{m}}}{k_{\mm{B}}T_{\mm{m}}}\frac{T-T_{\mm{m}}}{T}
\Big)-1\Big], {\forall T>T_{\mm{m}}}$,
where $a_{\mm{U}}$, $b_{\mm{U}}$ are material constants and $\Delta H_{\mm{m}}$ is the melting enthalpy.
 Under isothermal conditions, $U_{\mm{front}}$ keeps constant. Provided initial condition
 $d(0;D)=D$, we obtain
\begin{equation}\label{eq:d-t}
 d(t;D)=\left\{
 \begin{split}
 &D-2U_{\mm{front}}t,\  \forall\ t<D/2U_{\mm{front}};
 \\
 &0, \ \forall\ t\geq D/2U_{\mm{front}} .
\end{split}
\right.
\end{equation}
For the homogeneous melting in the inner region, we adopt the widely accepted KJMA  model\cite{Kooi2004Monte,Farjas2006Modification,Tomellini2014Kolmogorov}:
\begin{equation}\label{eq:chi-t}
\chi(t)=1-\exp\Big(-\mathbb{K}^4t^4\Big)
\end{equation}
where $\mathbb{K} = \Big(({\pi}/{3})I_{\mathrm{hom}}U^3\Big)^{1/4}$ is referred to as the overall KJMA rate constant, $I_{\mathrm{hom}}$ is the homogeneous nucleation rate and $U$ is the
growth velocity.
 It has been shown that liquid-solid interfaces formed through randomly nucleation in crystals propagate approximately like a macroscopically flat liquid-solid interface \cite{Forsblom2005Homogeneous}.
Thus, we assume $U=U_{\mm{front}}$.
According to the classical nucleation theory, the homogeneous nucleation rate is
$I_{\mm{hom}} = {k_{\mathrm{B}}T}/{\hbar V_{\mm{atom}}}\exp\Big(-{Q}/{k_{\mathrm{B}}T}-{16\pi\gamma_{\mm{sl}}^3}/{3k_{\mathrm{B}}T\cdot\big(\Delta H_{\mm{m}}\cdot(1-T/T_{\mm{m}})+\Delta E\big)^2}\Big)$ \cite{Mei2007Melting,Lu1998Homogeneous},
where $k_{\mathrm{B}}$ is the Boltzmann's constant, $\hbar$ is the Planck's constant,
$\gamma_{sl}$ is the solid-liquid interface energy, $V_{\mm{atom}}$ is volume per atom, $T_{\mm{m}}$ is the equilibrium melting
temperature, $\Delta H_{\mm{m}}$ is the fusion enthalpy change,  $Q$ is the activation energy for atomic diffusion
in the crystal lattice, $\Delta E=18\mu K\varepsilon_{\mm{m}}^2/(4\mu+3K)$ is the strain energy density related to the hydrostatic
eigenstrain $\varepsilon_{\mm{m}}$ corresponding to volume change during
melting, $\mu$ is the shear modulus,
and $K$ is the bulk modulus.
By taking Eq.\,(\ref{eq:d-t}) and Eq.\,(\ref{eq:chi-t}) into Eq.\,(\ref{eq:CHH}),
we obtain
\begin{equation}\label{eq:eta-t}
 \eta(t;D) = \left\{
 \begin{split}
 & 1 - \frac{(D-2U_{\mm{front}}t)^3}{D^3}\exp\Big(-\mathbb{K}^4t^4\Big),\
 \\ & \ \ \ \ \ \ \ \ \  \ \forall t<D/2U_{\mm{front}};
\\ & 1, \ \ \forall t\geq D/2U_{\mm{front}}.
\end{split}
\right.
\end{equation}
Eq.\,(\ref{eq:eta-t}) fully defines the melting kinetics of an isolated crystal involving both homogeneous and heterogeneous melting
mechanisms.

 The overall melting kinetics of plocrystalline aggregates depends on the grain-size-distribution (GSD).
Here, we use the Weibull distribution to describe the GSD \cite{Wang2003Stability,Fayad1999Steady}.
The Weibull probability density function is, \cite{Papoulis2002Probability}
\begin{equation}
f(D)=\frac{k}{\lambda}\Big(\frac{D}{\lambda}\Big)^{k-1}\exp\Big(-(D/\lambda)^k\Big), \forall D\geq0.
\end{equation}
where $k$ is known as the shape parameter and $\lambda$ is the scale parameter which is related to
 the mean-grain-diameter (MGD) $\bar{D}$ through $\lambda=\bar{D}/\Gamma(1+1/k)$ where $\Gamma$ is the gamma function.
And the $n\mm{th}$ raw moment of the Weibull distribution is $M_n = \int\limits_{0}^{+\infty} f(D)D^n\,\mm{d}D = \lambda^n\Gamma\Big(1+\frac{n}{k}\Big)$.
Based on the Weibull GSD, the overall transient liquid fraction $\bar{\zeta}(t)$ is
\begin{equation}\label{eq:CHH-Weibull}
\begin{split}
{\zeta}(t) =\frac{\int\limits_{0}^{+\infty} f(D)D^3\eta(t;D)\,\mm{d}D}
{\int\limits_{0}^{+\infty} f(D)D^3\,\mm{d}D} = 1-\frac{\exp\Big(-\mathbb{K}^4t^4\Big)}{M_3}
\times\int\limits_{2U_{\mm{front}}t}^{+\infty}
f(D)(D-2U_{\mm{front}}t)^3\,\mm{d}D
\end{split}
\end{equation}
Eq.\,(\ref{eq:CHH-Weibull}) defines the melting kinetics model for polycrystalline materials which explicitly couples the homogeneous and heterogeneous melting mechanisms,
with GSD directly incorporated.
We refer this melting kinetics model as the coupled-homogeneous-heterogeneous (CHH) model for polycrystalline materials.
It can be proved that $\lim\limits_{\bar{D}\rightarrow +\infty}\zeta(t) = \lim\limits_{\lambda\rightarrow +\infty} {\zeta}(t) =  1-\exp\Big(-\mathbb{K}^4t^4\Big)\equiv \zeta_{\mm{KJMA}}(t)$. Thus, in the limit of infinite mean grain size, the CHH model exactly reduces to the KJMA model. The CHH model is more general than the KJMA model and remains applicable for finite-grained materials where heterogeneous melting is more important.
 On the other hand, if homogeneous melting
is artificially suppressed by setting $\mathbb{K}=0$, then the CHH model reduces
to a pure heterogeneous melting kinetics model (the HET model): $\zeta_{\mm{HET}}(t)=1-1/{M_3}\times\int\limits_{2U_{\mm{front}}t}^{+\infty}f(D)(D-2U_{\mm{front}}t)^3\,\mm{d}D$.

\begin{figure}[!htbp]
\centering
\includegraphics[width=0.5\textwidth]{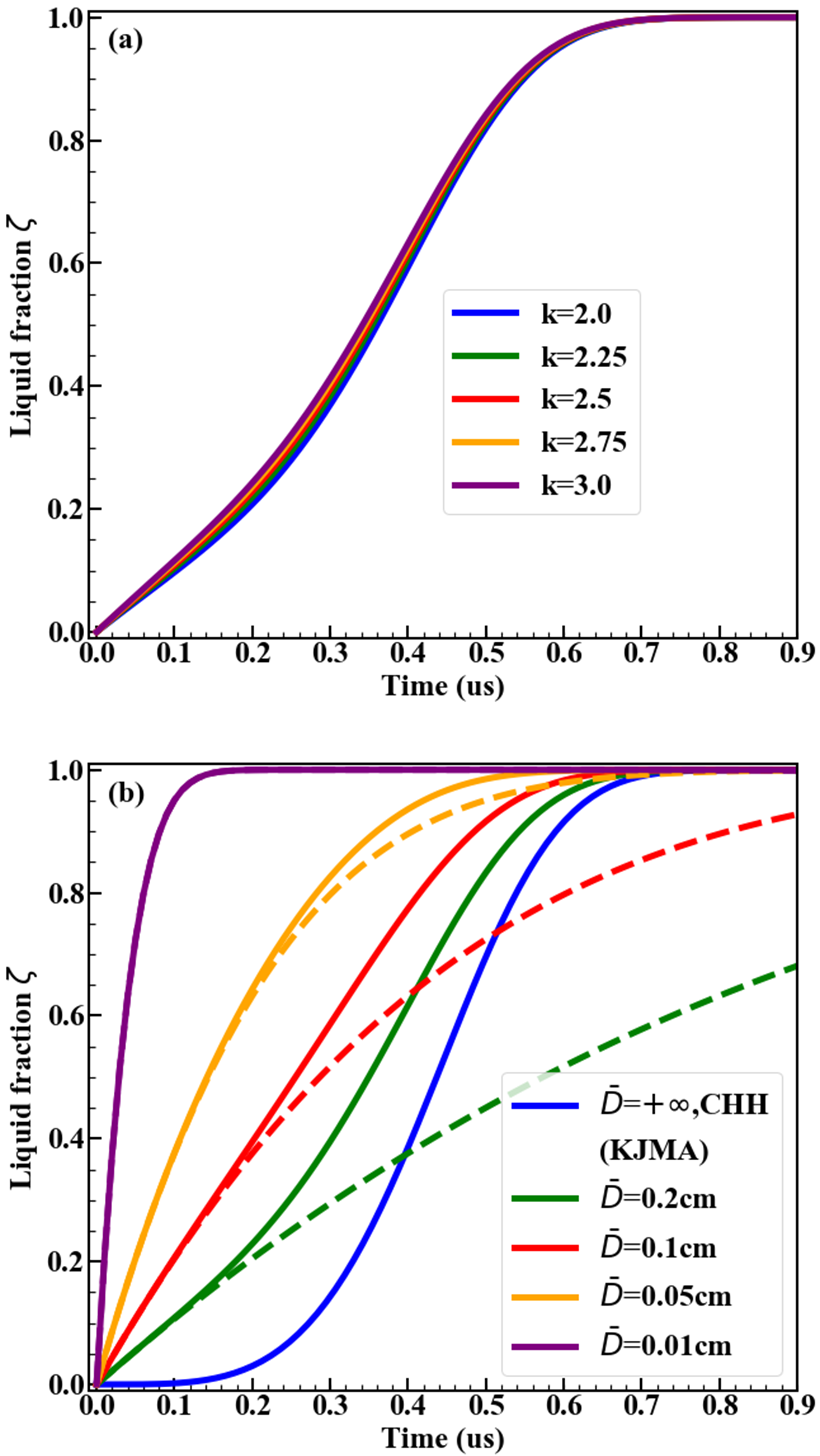}
\caption{\label{fig:ZetaVsTime} Effects of GSD on melting kinetics under constant temperature $T=1128\,\mm{K}$. (a) Plots for fixed $\bar{D}=0.1\,\mm{cm}$ and various $k$;
(b)  Plots for fixed $k=2.5$ and various $\bar{D}$.
In (b), the solid lines are predictions of the CHH model;
The dashed lines are predictions of the HET model. In the case of
$\bar{D}=+\infty$, the CHH model is equivalent to the KJMA model.
}
\end{figure}

\begin{table}[!h]
\centering
\begin{threeparttable}
\centering \caption{\label{tab:parameters} Material parameters for melting kinetics of polycrystalline Al.}
\begin{tabular}{ c c  c c  c c c c c}
 \hlinew{1.3pt}
  $a_{\mm{U}}$ & $b_{\mm{U}}$ & $\varepsilon_{\mm{m}}$ & $\mu$(GPa) & $K$(GPa) & $Q$(ev) & $\Delta H_{\mm{m}}(\mm{kJ/mol})$
  & $\gamma_{\mm{sl}}(\mm{J/m^2})$ & $T_{\mm{m}}$(K)
\\\hline
  0.206 & 5.28 & 0.02 & 26 & 76 & 1.48 & 10.7 & 0.108 & 933
\\
\hlinew{1.3pt}
\end{tabular}
\end{threeparttable}
\end{table}

\section{Applications}
We apply the model to study melting process of superheated
 polycrystalline Al (pc-Al).
 The parameters for the model are provided in Table.\,\ref{tab:parameters}.
 The parameters $a_{\mm{U}}$ and $b_{\mm{U}}$ are
 taken from Ref.\cite{Mazhukin2016Temperature} and other parameters
are taken from Ref.\cite{Kelton1991Crystal}.
 In calculations, the infinite integral in Eq.\,(\ref{eq:CHH-Weibull})is approximated by $\int\limits_{2U_{\mm{front}}t}^{+\infty}f(D)(D-2U_{\mm{front}}t)^3\,\mm{d}D \approx \int\limits_{2U_{\mm{front}}t}^{D_{0.99999}}f(D)(D-2U_{\mm{front}}t)^3\,\mm{d}D$  where $D_{0.99999}$ is the 0.99999 quantile of the Weibull distribution. The finite integral is then calculated numerically by cumulative trapezoid integral method.

 Fig.\,\ref{fig:ZetaVsTime} displays the liquid
 fraction versus  time ($\zeta-t$) plots under constant temperature $T=1128\,\mm{K}$ for pc-Al,
  revealing the effects of GSD on melting kinetics.
  The shape parameter $k$ in the Weibull GSD depends on specific generation methods for polycrystalline materials.
Generally, $2<k<3$ \cite{Wang2003Stability}.
From Fig.\,\ref{fig:ZetaVsTime}(a), melting kinetics
 is insensitive to the shape parameter $k$. Therefore, in following analysis, we fix $k=2.5$ and focus on the effects of
 MGD, i.e., $\bar{D}$.
 When $\bar{D}=+\infty$,  the CHH model predicts the same curve as the KJMA model.
 In this case, the whole melting process include three stages. In the initial stage, the amount of growing nuclei is small and the  liquid fraction increases slowly.  The melting rate (the slope of the curve) increases as the nuclei number increases.
 In the latter stage, the melting rate slows down because of lack of mother solid phase.
 This melting process can be summarized as a "slow-fast-slow" three-stage process.
 Characteristics of melting in polycrystalline materials with finite grain size depends on $\bar{D}$.
 From Fig.\,\ref{fig:ZetaVsTime}(b), coarse-grained materials (e.g., $\bar{D}=0.2\,\mm{cm}$),
 would experience the similar "slow-fast-slow" melting stages as predicted by the KJMA model.
 Both the CHH model and HET model predict faster melting for finer-grained materials.
 In cases of $\bar{D} = 0.2\,\mm{cm}$ and $0.1\,\mm{cm}$, the $\zeta-t$ curve of the CHH model deviates heavily from that predicted by the HET model. As the MGD decreases, it gets closer to the HET model,
 indicating that heterogeneous melting plays more important roles in finer-grained materials.
 During heterogeneous melting, melting rate is proportion to the product of the area of the melting front surface and its motion velocity. The diameter of the spherical melting front steadily decreases. As a result, heterogeneous melting rate reaches its maximum at the very beginning and then monotonically decreases as melting progresses, different from the "slow-fast-slow" three-stage melting history predicted by KJMA model.

\begin{figure}[!htbp]
\centering
\includegraphics[width=0.5\textwidth]{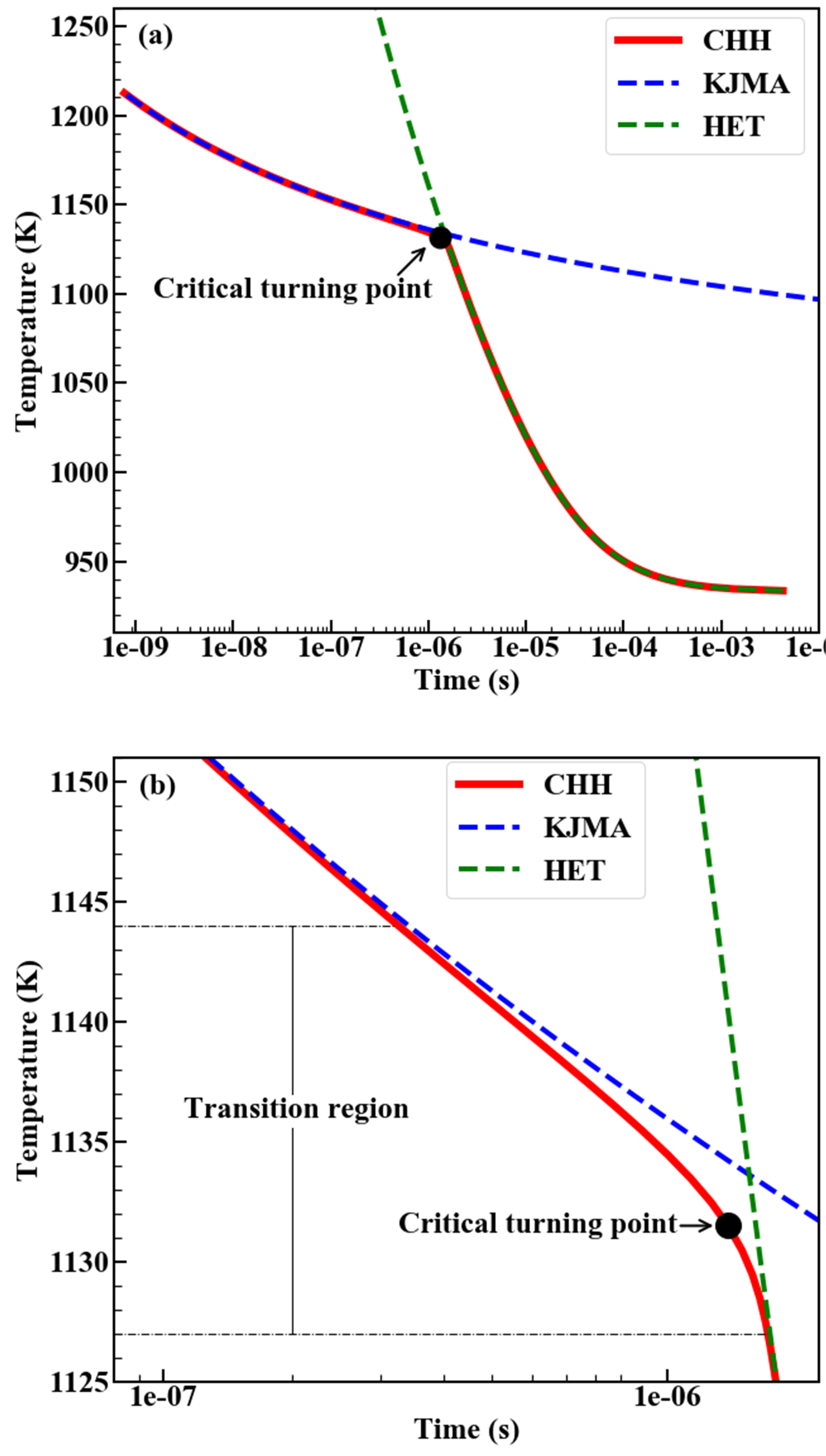}
\caption{\label{fig:TTT} (a) The TTT diagram for polycrystalline Al with MGD $\bar{D}=0.1\,\mm{cm}$;
(b) An enlarged view of the part near the critical turning point.
}
\end{figure}

Temperature-time-transformation (TTT) diagram is an efficient approach to characterize melting kinetics in the
  whole superheating regime.
 Fig.\,\ref{fig:TTT}(a) compares the the TTT diagram (plots of temperature versus time corresponding to $\zeta=99\%$
 for pc-Al with $\bar{D}=0.1\,\mm{cm}$,
 in comparisons to predictions of the KJMA
 and HET models.
The TTT curve for
melting of pc-Al differ fundamentally from those predicted
by the KJMA and HET models.
 The TTT curves based on the KJMA and HET models
 are smooth and globally convex.
However, the CHH-based TTT curve is zigzag and nonconvex.
There is a sharp turning which divide the CHH-based TTT curve into two major branches. The lower branch well coincides with the curve of the HET model
while the upper branch well coincides with the curve of the KJMA model. This indicates that
heterogeneous melting make the overwhelming contributions below the turning part while homogeneous melting paly the overwhelming role
above the turning part.
Fig.\,\ref{fig:TTT}(b)
shows an enlarged view of the turning part. From Fig.\,\ref{fig:TTT}(b), the turning part obviously deviates from both the KJMA-based
and the HET-based TTT curves, indicating a transition region where niether
the heterogeneous melting nor the homogeneous melting plays the overwhelming role. Instead, the contributions of the two melting mechanisms are comparable .


The transition part is very short comparing to the whole TTT curve.
It is corresponding to a very narrow transition temperature range in which the two melting mechanisms make comparable
contributions.
When temperature is in this transition range, contributions of the two melting mechanisms are comparable.
Homogeneous melting dominates if temperature is above the transition range while heterogeneous melting
 dominates if temperature is below the transition range.
To quantitatively probe the competitions between
the two melting mechanisms, we focus on the critical point where the curvature reaches the maximum
on the TTT curve, marked by the filled circle in Fig.\,\ref{fig:TTT}.
 The tangent of the curve turns most sharply at this point, which indicates obvious switching of the primary
 mechanism across the point. The temperature corresponding to this point is referred to as the critical turning temperature
  and denoted as $T^\#$.
For pc-Al with $\bar{D}=0.1\,\mm{cm}$, it is calculated that $T^\# = 1131K$. 


\begin{figure}[!htbp]
\centering
\includegraphics[width=0.5\textwidth]{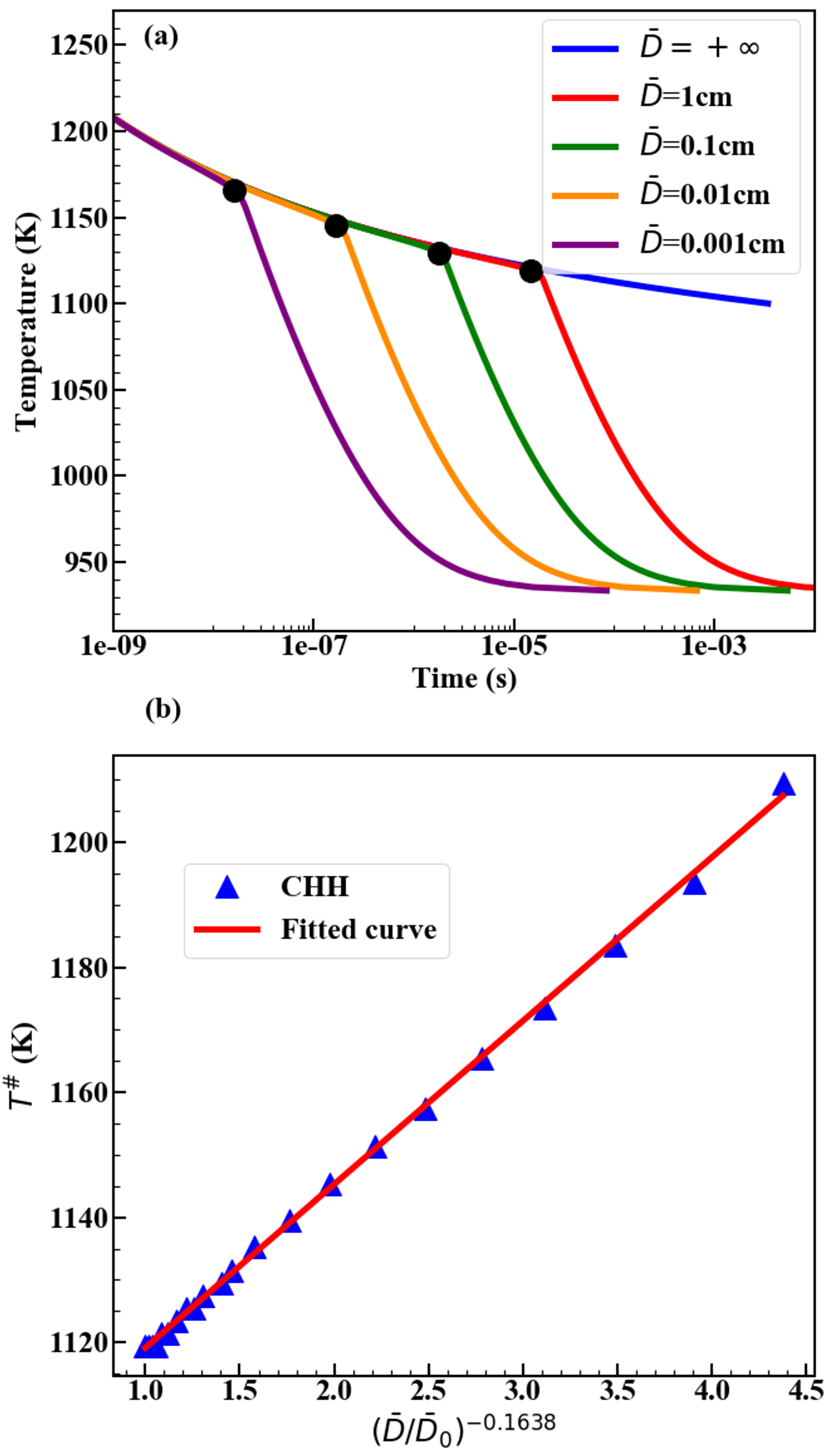}
\caption{\label{fig:MGD-effects} (a) The TTT diagrams for polycrystalline Al with different MGDs;
(b) The MGD dependence of the critical temperature, where the solid line is fitted curve of the CHH predictions through the power law.
}
\end{figure}
Fig.\,\ref{fig:MGD-effects}(a) compares the TTT curves of pc-Al with different MGDs.
It is shown that the  the turning part and the critical temperature $T^\#$ raises as the MGD decreases,
 which indicates that heterogeneous melting would play the dominant role in a wider superheating temperature range for finer-grained materials.
This can be explained by analyzing competitions between the two melting mechanisms involving both MGD effects and temperature effects.
 On one hand, increasing temperature would enhance the contributions of homogeneous melting and weaken the contributions of heterogeneous melting. This is because the homogeneous nucleation rate exponentially increases as temperature increases \cite{Lu1998Homogeneous}, whereas motion velocity
of the heterogeneous melting front is much less sensitive to the temperature \cite{Ivanov2007Kinetic}.
On the other hand, the heterogeneous melting rate increases as MGD decreases.
As a result, a higher superheating temperature are required to make
contributions of homogeneous melting overweigh that of heterogeneous melting in finer-grained materials.
This explains why the turning part and $T^\#$ increases as the MGD decreases.
Fig.\,\ref{fig:MGD-effects}(b) displays  $T^\#$ as a function of MGD ranging from
several nanometers to several millimeters.
In this range, the dependence of the critical temperature $T^\#$ on $\bar{D}$ is well-fitted by a power law:
$T^\# = a+b\Big(\frac{\bar{D}}{\bar{D}_0}\Big)^c$
where $\bar{D}_0=1\,\mm{cm}$ is a reference MGD, $a$, $b$, $c$ are fitting parameters.
For Al, the fitting parameters are $a=1093\,\mm{K}$, $b=26.2\,\mm{K}$, $c=-0.1638$, as shown in Fig.\,\ref{fig:MGD-effects}(b).

Experimentally, it is rather difficult to quantitatively distinguish  the melting processes inside  polycrystals. In limited experiments \cite{Luo2003Superheating,Luo2003Maximum,Rethfeld2002Ultrafast}, homogeneous melting
was considered as the major melting mechanism at high superheatings under ultrafast heating while heterogeneous melting was considered as the major melting
mechanism at low superheatings. These results are also supported by computer simulations\cite{Ivanov2007Kinetic}. The predictions of the present model are qualitatively consistent with these reults.

\section{Summary and discussions}

In summary, a new kinetics
model which explicitly coupling the two melting mechanisms (homogeneous and heterogeneous melting) is proposed to better understand melting kinetics of polycrystalline materials.
Through the model, it is possible to quantitatively analyze the
 effects of grain-size-distribution on the  melting kinetics  as well as
 competitions between the two melting mechanisms.
 It is found that melting kinetics is nonsensitive to the shape parameter of the Weibull grain-size-distribution but is strongly
 dependent on the mean-grain-diameter.
The coupled-homogeneous-heterogeneous model predicts  nonconvex TTT diagrams with a three-part structure:
 a lower part which is dominated by heterogeneous melting, a upper part which
is dominated by homogeneous melting and a narrow transition part where the two melting  mechanisms make comparable contributions.
It allows to determine the critical temperature at which the primary melting mechanism switches.
The critical temperature increases as the mean-grain-diameter decreases following a negative power law.

Due to the extreme complexity of the problem, there  are several limitations of the model. Firstly, our study hasn't considered effects of defects inside the
crystals (e.g., point defects, dislocations and vacancies).
Presence of these defects would generally facilitate liquid nucleation
inside the crystals and thus enhance the contributions of homogeneous melting.
The model also
hasn't taken into the boundary effects on the accuracy
of the KJMA model which, in principle, was developed for infinite medium \cite{Tomellini2016Modeling}.
In addition, the results are confined to the Weibull grain-size distribution.
Moreover, the present work only focus on isothermal transforma-
tion analysis. Isothermal transformation analysis is impor-
tant for understanding melting kinetics under ultafast heating
rate. Experimentally, ultafast heating rate ($10^12 \mm{K/s}$) can be
achieved through shock-wave loading and intense laser irradia-
tion, which may result in superheating as high as $0.5\,T_{m}$
\cite{Luo2003Superheating,Luo2003Maximum,Luo2004Shock,
Rethfeld2002Ultrafast}.
Heating rate effects, which is important for low heating rate
conditions, is not involved here.
However, the present work, for the first time, combine the two most important issues (i.e., random nucleation and growth in crystal
   interiors and heterogeneous melting from grain boundaries.) and can be a good starting point for further research.


This work is supported by the National Natural Science Foundation of China (No.11772068), the Foundation of LCP and
the	Presidential Foundation of China Academy of Engineering Physics (No.YZJJLX2017011).


%

\end{document}